# RELATION BETWEEN WIND AND WIND CURRENTS ON THE SYNOPTIC SCALE WITH ACCOUNT OF WAVE CONDITIONS

Polnikov VG,

Obukhov Institute of Atmospheric Physics. 119017 Moscow, Pyzhevsky, 3; polnikov@mail.ru

Kabatchenko IM,

Zubov State Oceanographic Institute, 119034 Moscow, Kropotkin lane., 6; wavelab@yandex.ru

**Abstract**

A version of model is proposed, which is aimed for getting parameters of the atmospheric layer and upper water layer with account of the wind-wave state. The dynamics of the atmospheric boundary layer is realized in version of papers [1, 2], and the dynamics of the upper layer is realized in the framework of Ekman layers in the atmosphere and the sea [3]. In the latter case, the Kitaigorodskii [4, 5] approach was used for describing the dynamics of the air-water interface. The key parameters of the atmospheric layer and upper water layer (the friction velocity and the speed of drift currents) are calculated for typical wind-wave situations. Satisfactory quantitative agreement between calculated and traditionally observed values is shown.

## 1. Introduction

It is well established that the wind waves play an important modulating role in the process of large-scale air-sea interaction [6, 7]. In this connection, the problem is arising to construct a unified model describing the dynamics of the air-water interface, acceptable on a wide range of scales: from hundreds of meters and seconds to the global scale (thousands of miles and hours). To solve this problem, we are basing on studies showing that the link among different scales and processes at the air-sea interface can be realized by a numerical spectral model of wind waves [1, 2, 6, 8, 9].

A general approach to this issue was described in a recent paper [6] where the processes of momentum and energy fluxes from wind to waves, and then to the upper layer of the sea were reviewed, analyzed, and partly calculated. It was shown that the combined (large scale) model of the air-water interface should include, first of all, a spectral model of wind waves. Additionally, this model should be added by 2 blocks: the block of dynamic atmospheric boundary layer (DBL) and the block of dynamic upper water layer (DUL). Wind-wave model (which includes three evolution mechanisms: the mechanism of energy transfer from wind to waves, IN, mechanism of nonlinear redistribution of wave energy, NL, and mechanism of wave energy losses, DIS) is responsible for the dynamics of two-dimensional wave spectrum, $S(\mathbf{k}, \mathbf{x}, t)$ given in the wave number space, $\mathbf{k}$, and in the space of coordinates, $\mathbf{x}$, and time, $t$. The block of DBL is responsible for matching the wave parameters with the parameters of the atmospheric layer (for example, matching wave spectrum $S$ and friction velocity $U_*$). The block of DUL plays the same role but for matching the wave parameters with the parameters of the upper water layer (eg, matching wave spectrum $S$ to the coefficient of vertical exchange $K_T$ and drift velocity $V_d$) (for details, see example in [8]). The united model of such type is proposed to be classified as a wind-wave model of the fifth generation[6], because it is conceptually different from the models of previous the third and fourth generations that do not affect the dynamics of the upper water layer of the sea(see examples of the models in [7]. A block diagram of this model is shown in Fig. 1.

Analysis of previous studies [1, 6, 8, 9] have shown that the most poorly designed part of the fifth generation of models is the block of DUL. This work gives a possible version of solving the issue.

## 2. Pose of general problem

Let us consider two interacting boundary layer: atmospheric and upper water. The main external parameter driving these layers is the local geostrophic wind $U_g$. Herewith, as noted in the introduction, there is an additional, internal parameter mediating interaction of atmospheric and marine layers. This internal parameter is related with the state of the sea surface which has a structure of the wavy interface. Due to the statistical feature of the interface, the most general representation of its state is given by the two-dimensional wave-vector spectrum $S(\mathbf{k})$ or by its equivalent in the frequency-angular representation $S(\omega,\theta)$ (for details see [6]).

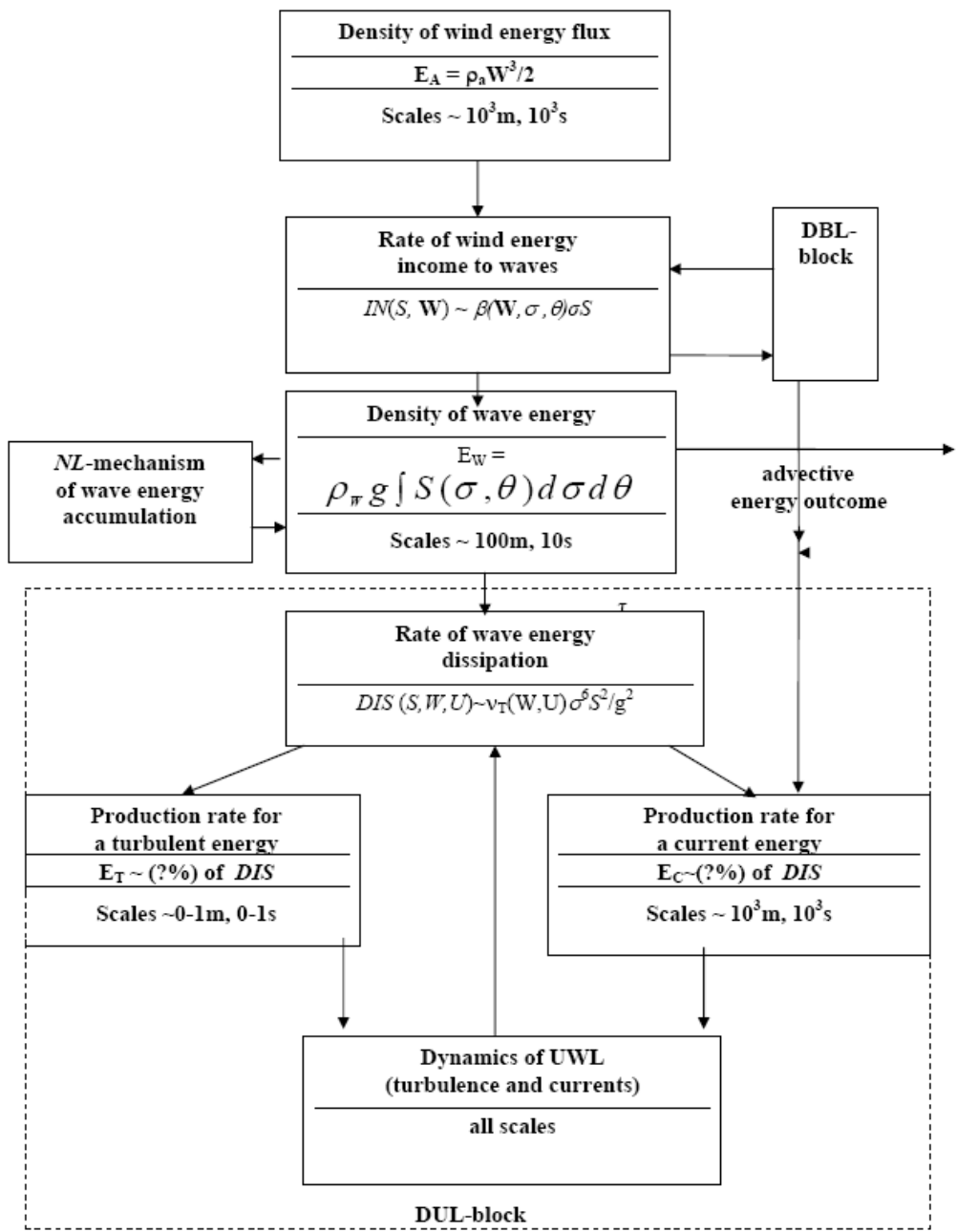

Fig. 1. General scheme of the wind wave model of the fifth generation, showing the path of momentum and energy exchange at the air-sea interface (following to [6]).

The value geostrophic wind velocity $U_g$, driving dynamic characteristics of the atmosphere layer, can be calculated from by the model [10]:

$$-\frac{1}{\rho_a}\frac{\partial p}{\partial n} + f_k U_g \pm \frac{U_g^2}{R} = 0, \qquad (1)$$

where $p$ is the atmospheric pressure field in the planetary surface layer, $R$ is the radius of the isobars curvature, $\rho_a$ is the air density. Typically, the surface pressure field is available in graphical form. Usually, this is the ring weather maps. In any case, we suppose that the pressure field (and the wind field, consequently) is available at the numerical grid points.

To calculate the spectrum of waves, one should use one or another numerical wind-wave model, a concrete representation of which is not principal here (see, for example, [2, 6, 7]). However, for the model use, it is required the wind at the standard horizon, namely, at the height $z = 10$ m, i.e. the wind $U_{10}$. To calculate it, one needs a specific model of global atmospheric boundary layer. In turn, the latter model requires knowledge of the surface roughness parameters which values are determined by the wind-wave spectrum. Consequently, to calculate the characteristics of wind speed and condition of the underlying surface atmospheric layer, it needs a selfconsistent model, which is called as a block of dynamic boundary layer (DBL). One variant of this model, which is used in this paper, was presented in [1].

Further, the value of surface roughness of the interface allows to introduce the concept of momentum flux in the upper water layer, and make an analogy between the dynamics of atmospheric boundary layer and dynamics of upper water layer by using estimates of the Ekman layer for ocean currents, following Kitaigorodskii[5]. This is one of the easiest ways to determine such characteristic of the DUL as the drift current, $V_d$.

As far as the characteristics of DBL and DUL are uniquely determined and dependent on the wave state, in the above approach one can assume that the task of matching dynamics for DBL and DUL via the wavy state of the interface is solved. Implementation of the proposed approach, its details, and some results of calculations are shown below.

### 3. Model of dynamic boundary layer

The total wind stress at the sea surface, $\tau_a$, can be represented by two terms as follows:

$$\tau_a = -\rho_a U_*^2 = \tau_t + \tau_w \qquad (2)$$

where $\tau_t$ is the momentum flux at the interface in the absence of waves, and $\tau_w$ the change in momentum flux, provided by presence of waves and corresponding to the energy transfer from wind to waves.

We assume that the flow $\tau_t$ corresponds to the turbulent air flow over a tight and smooth underlying surface. Roughness parameter for such flow is known [11], and given by

$$z_0 = z_{0v} = a_v \nu / U_{*v} \qquad (3)$$

In this relation, $a_v \cong 0.1$ and $\nu$ is the kinematic viscosity of air, $U_{*v}$ is the friction velocity of a flow over a aerodynamically smooth surface. With the known value of the geostrophic wind, for the connection speeds $U_{*v}$ and $U_g$ can use the Kazansky - Monin resistance law [4]

$$U_{*v} = \kappa U_g \left[ \left( \ln \frac{U_{*v}}{f z_{ov}} - B(\mu) \right)^2 + A^2(\mu) \right]^{-1/2} \qquad (4)$$

where $f$ is the Coriolis parameter ($f = 2\Omega \sin\varphi$, $\Omega$ is the angular velocity of the Earth rotation, $\varphi$ is the local latitude), parameters $A$ and $B$ are determined by the thermal stratification of the atmospheric layer, and $\mu = h/L$ ($h$ is thickness of the atmospheric boundary layer, $L$ is the Monin - Obukhov length scale[12], which is assumed to be known). The angle of deviation of surface wind from the direction of the geostrophic wind, $\alpha$, as determined by the parameter of thermal stratification [12]: $\sin\alpha = -\dfrac{U_*}{\kappa U_g} A(\mu)$. Thus, relations (3) and (4) completely determine the value of $\tau_t = -\rho_a U_{*v}^2$.

On the other hand, the momentum flux to waves $\tau_w$ is determined by the two-dimensional frequency-angular spectrum via the so-called wind-to-waves pumping function which describes the mechanism of energy transfer from wind to waves. Generally accepted expression for $\tau_w$ has the form [6, 13]

$$\tau_w(z=0) \equiv \tau_w(0) = \rho_w g \int \frac{k \cos(\theta)}{\omega} IN(S, \mathbf{W}, \omega, \theta) d\omega d\theta . \qquad (5)$$

Substitution of representation $S(\omega,\theta) = S(\omega)\Psi(\omega,\theta)$ (where $S(\sigma)$ is a one-dimensional wave frequency spectrum, and $\Psi(\sigma,\theta)$ its frequency-angular distribution) and the well-known representation for the pumping function [6, 7] of the form

$$IN(..) = \left\{\frac{\rho_a}{\rho_w} \beta(U_*, \omega, \theta)\right\} \omega S(\omega, \theta) \qquad (6)$$

(where $\beta$ is the known wave increment coefficient, and $\rho_w$ is the water density), yields:

$$\tau_w = -\rho_a \int_0^{\omega_+} S(\omega) \omega^2 d\omega \int_0^{2\pi} \beta(U_*, \omega, \theta) \Psi(\omega,\theta) \cos(\theta) d\theta. \qquad (7)$$

Here: $\omega_+$ is the upper limit of frequency range, where the waves no longer interact with the wind (at these frequencies, the height of wind waves are so small that they do not rise above the
5

viscous sublayer). In practical calculations one can use the following empirical approximation for coefficient $\beta$ [14]:

$$\beta = \left\{ \left[ b_1 (U_* k/\omega)^2 + b_2 (U_* k/\omega)^2 + b_3 \right] \cos(\theta) - b_4 \right\}, \quad (8)$$

where $b_1 = 4*10^{-2}$; $b_2 = 5,44*10^{-3}$; $b_3 = 5,5*10^{-5}$; $b_4 = 3,1*10^{-4}$.

With the parameterizations given above, formula (2) can be rewritten as

$$U_*^2 = U_{*v}^2 + C_1 U_*^2 + C_2 U_* + C_3, \quad (9)$$

where the value for $U_{*v}$ follows from (4), and the coefficients $C_1 - C_3$, following from (5) - (8), are the functions of the wave spectrum shape.

To ensure the integration in (7), we use

$$\omega_+ = (gk_0)^{1/2}, \; k_0 = \alpha_0 U_{*t}/\nu, \; \alpha_0 \cong 10^{-2}. \quad (10)$$

Rations (10) restrict the momentum flow to the small-scale component of wind waves in the viscous sublayer (thickness $h_v \cong 5\nu/U_{*t}$), which, on the physics, does not contribute to the aerodynamic resistance of the rough surface.

Thus, the task of determining characteristics of DUL is reduced to solving a quadratic equation (9) with respect to $U_*$. To obtain the real (effective) roughness parameter $z_0$, including the part generated by waves, we can use the same formula of the Kazansky - Monin resistance law (4). As a result, one gets

$$z_0 = \frac{U_*}{f \exp\left( \sqrt{\frac{\kappa U_g}{U_*} - A^2(\mu)} + B(\mu) \right)} \quad (11)$$

The parameter $z_0$ obtained by this procedure was compared with the known field experiment [15]. The corresponding calculations were performed for different values of the geostrophic velocity $U_g$ and different values of the wave age $A$, determined by ration $A = C_p / U_*$, where $C_p$ is the phase velocity of waves at the peak-of-spectrum frequency. Herewith, in order to compare calculations with field data [15], the dependence of $z_0$ on the inverse wave age is used.

At this stage of investigation, the wave spectrum used in the calculation was specified by the standard form of the JONSWAP spectrum[7] with a changing value of the frequency peak $\sigma_p$:

$$S(\omega) = \frac{\alpha g^2}{\omega^5} \exp\left[ -\frac{5}{4} \left( \frac{\sigma_p}{\omega} \right)^4 \right] \gamma^G, \text{ with } G = \exp\left[ -\frac{1}{2} \left( \frac{\omega - \sigma_p}{\Delta \sigma} \right)^2 \right], \quad (12)$$

where $\Delta\sigma = \begin{cases} 0,07, & \omega/\sigma_p \leq 1 \\ 0,09, & \omega/\sigma_p > 1 \end{cases}$, and parameters $\alpha$, $\sigma_p$, $\gamma$ are the functions of the dimensionless acceleration [1]. Angular spreading was used a simple frequency-angular form

$$\Psi(\omega,\theta) = \frac{2}{\sqrt{\pi}} \cos^2(\theta - \theta_w), \qquad (13)$$

with a constant value of wind direction $\theta_w$.

Figure 2 shows the dimensionless roughness parameter against the inverse wave age. From the figure we can draw the following conclusions:

1) The dimensionless roughness parameter depends strongly both on the external parameter: geostrophic wind velocity $U_g$, and on the internal, wave state parameter: in this case, inverse wave age, $A^{-1} = U_*/C_p$.

2) The calculation results obtained in the model DBL have a quite satisfactory agreement with observations [15].

The latter conclusion gives a final estimate of the proposed DBL model quality.

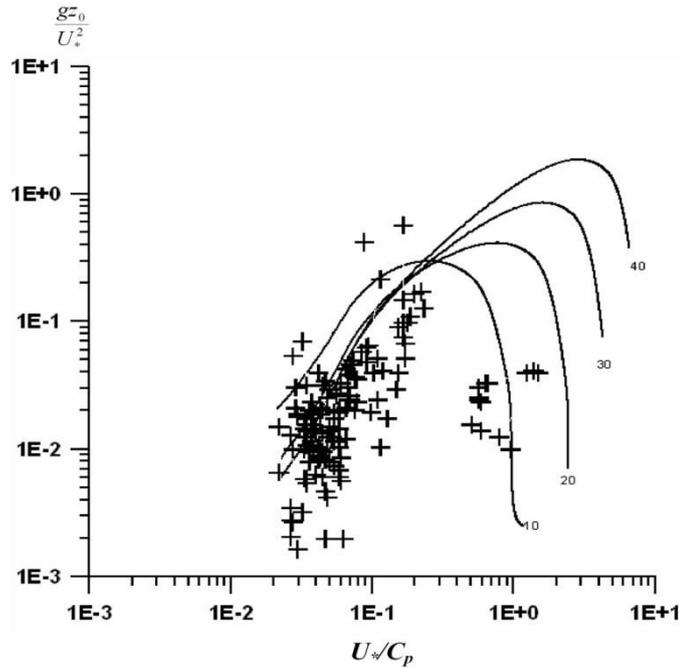

Fig. 2. Dimensionless roughness against the inverse wave age (points are the experimental data [15]), solid lines are calculation by the model for different geostrophic wind speeds $U_g$ (m / s).

## 4. Dynamic upper layer model

Marine boundary layer located in the vicinity of the interface is strongly turbulized [5]. This is due to several factors. First of all, turbulence is provided by the waves overturning and related processes such as: the white capping, air bubbles, water dripping, and so on. Secondly, the surface layer of water evaporates, which leads to the formation of a water layer with higher salinity. Sinkable salt water vigorously stirs a surface layer. In addition, the turbulence of the surface water layer is supported by the process of bubbles floating. These bubbles are formed during overturning of the wave crests when tipping the air trapped with them into the water.

All these complicated processes make it impossible to describe the marine boundary layer in details as well as atmospheric one. We should mention that there is an ideology [6] to use the dissipation function of wave energy, *DIS(S)*, for the turbulent describing the total effect of these processes. However, at meanwhile this ideology is not detailed to the level of its use in numerical calculations. Therefore, at this stage of building a pioneer model matching DBL and DUL, we will involve more simple considerations, the fairness which will be estimated a posteriori.

Traditionally, the dynamical characteristics of the marine boundary layer are described in terms of total current flows $\mathbf{T}(T_x, T_y)$ [3]:

$$T_x = \int_d^0 U(z)dz, \quad T_y = \int_d^0 V(z)dz, \tag{14}$$

where the *x*-coordinate coincides with the direction of the wind on the surface, and *y* is perpendicular to it, *d* is the thickness of the marine boundary layer. In this case, according to [3], one can write

$$T_x = 0, \quad T_y = \frac{\gamma^2}{\Omega \sin \varphi}, \tag{15}$$

where

$$\gamma = \sqrt{\frac{\rho_a}{\rho_w}} U_*, \tag{16}$$

and the depth of friction layer is determined by the formula [3]:

$$d = \frac{\gamma}{\Omega \sin \varphi}. \tag{17}$$

Thus, the module average velocity of the Ekman flow is $V_E = \frac{|\mathbf{T}|}{d} = \sqrt{\frac{\rho_a}{\rho_w}} U_*$

The Ekman friction layer (17) can reach several hundred meters, but the average current speed in it is less than velocity scale variability of wind waves. For such scales the greatest interest has the drift velocity on the surface. In order to estimate its speed, in [2] it was used the equality of the shear stress in water and air at the interface.

The essence of the approach adopted in [2] is as follows. Let us use the notions introduced by Kitaigorodskii SA. Then, for the so-called friction velocity below the water surface, $V_s$, according to [5], one can write

$$V_s = U_* \sqrt{\frac{\rho_a}{\rho_w}} . \qquad (18)$$

Such ratio is acceptable for the case of a steady (fully developed) wave field, when all momentum flux from wind to waves goes to the upper water layer. In this case, passing to the system of coordinates moving opposite to the water drift with velocity $V_d$, in the new coordinates we obtain an analogue of the wind boundary layer located in the upper water.

Let us use now the Kazansky - Monin resistance law (4) for the mentioned boundary layer in the water. Assuming that far from mean water surface, the drift speed $V_d$ is the analog of geostrophic velocity $V_g$, and the flow velocity is zero at the interface (in the adopted coordinate system), according to formula similar to (4), we obtain the following expression for the drift current:

$$V_d = \frac{V_s}{\kappa} \left[ \left( \ln \frac{V_s}{f z_0} - B \right)^2 + A^2 \right]^{1/2} , \qquad (19)$$

where $z_0$ is a roughness parameter of the surface "from below". It is for its assessment, in fact, one should use the dissipation function *DIS*. However, this issue, as already noted, is not resolved yet. Therefore, to set $z_0$ we attract an empirical formula [16]:

$$z_0 \approx 0.1 * \frac{K_0}{V_s} \qquad (20)$$

where the function of the turbulent viscosity $K_0$ is ascribed by the recommended value $K_0 \approx 2 \times 10^{-4} * \frac{U_a}{g}$ (here, $U_a$ is the wind speed at the standard horizon, and g is acceleration due to gravity).

Next, to assess the value of $V_d$, we will use the so-called wind coefficient

$$\eta = V_d / U_a \qquad (21)$$

Coefficient $\eta$ calculated by formulas (18-20) for different geostrophic wind speeds is shown in Fig. 3, with its dependence on the inverse wave age.

Figure 3 shows that for the developed waves (the inverse wave age is about of 0.025) the drift has the order of 2% of wind speed $U_a$, regardless of geostrophic velocity. For young waves (inverse age is more than 0,1) the wind coefficient $\eta$ does increase with dependence on the geostrophic velocity: from value of 0.022, for geostrophic velocity of 10 m/s, to value of 0.043, for the geostrophic speed of 40 m/sec.

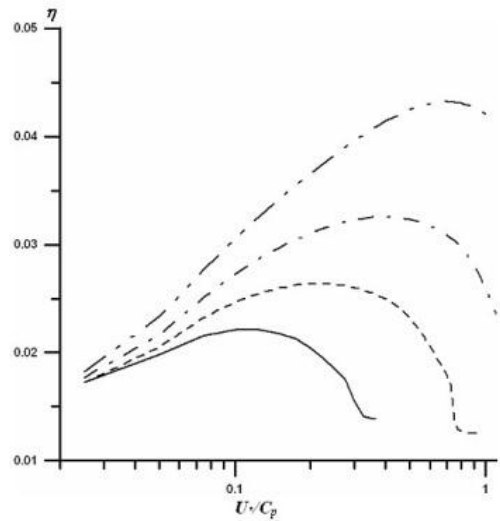

Fig. 3. Dependence of the wind coefficient $\eta$ ( the ratio of drift velocity on the surface to the wind speed at 10 m) against inverse wave age for different values of the geostrophic wind velocity: solid line -10 m / s, dotted line -20 m / s, dot-dash - 30 m / s, two dots-dash - 40 m / s.

The values of the wind coefficient obtained for steady waves agree rather well with the instrumental evaluations of this parameter, for example, in the Baltic Sea [17]. Additionally, the increase of wind coefficient for the young waves is confirmed in a number of numerical experiments (see, for example, [18]). This fact has a good physical basis: the young waves have a high roughness, what leads to a large momentum flux from wind, that does a strong drift.

Thus, the proposed dynamic model of upper water layer, in terms of determining the drift velocity as a function of the wave state, seems to be quite acceptable.

### 5. Conclusion

In conclusion we should note that the proposed version of matching the dynamic parameters of boundary layers of the atmosphere and the upper sea, including dependence on the state of unrest has the trial character. It can be used at present as the subjected of thorough verification started earlier while using the Russian coupled Atmospheric-Wave Model (RAWM) [2]. However, it should be noted that this goal will require rather complicated simultaneous measurement at the interface of the following parameters: the momentum flux from wind to waves $\tau_a$ (i.e, $U_*$), two-dimensional spectrum of emotion $S(\mathbf{k})$ (or $S(\sigma,\theta)$), and the drift velocity $V_d$. Moreover, such measurements should be performed for a wide variety of wind-wave situations. It is clear that possibility of realization of such measurements is difficult to forecast. Regarding to using the literature data to this aim, to some extent, this resource has already been used in this study and in dissertation [2]. But the further development of theoretical foundations for constructing the blocks of DBL and DUL has no obstacles mentioned above, and such work is planned.

*Acknowledgements.* We are grateful to MM Zaslavsky for numerous discussions of the issues involved. This work was supported by RFBR, grant № 08-05-13524-ofi-ts, 09-05-00857-a.

REFERENCES

1. Zaslavsky, MM, Zalesny VB Kabatchenko IM, Tamsalu R. Self-consistent description of the atmospheric boundary layer wind waves and sea currents. - Oceanology, 2006. , 46, № 2. S. 178-188
2. Kabatchenko IM Modeling wind waves. Numerical calculations for climate research and design of hydraulic structures - ABSTRACT of thesis for degree of Doctor of Geographical Sciences, 2007, JSC "Alliance Document Centre", 41.
3. Phillips OM Dynamics of upper ocean, $2^{nd}$ edn. Cambridge University Press, Cambridge. 1977. 336 pp
4. Kazansky, AB and Monin, AS On the dynamic interaction between the atmosphere and surface. - Math. Akad Geophys., 1961, N5, s.786-788
5. Kitaigorodsky SA Aerodynamic roughness of the sea surface above and below. - Publishing Center "Meteorology and Hydrology", M, 2002, 16.
6. Polnikov VG The role of wind waves in the dynamics of the interface water-air - Izv. RAS, Ser. FAO. 2009. T.45, № 3. p. 371-382.
7. Cavaleri L. et al (WISE group). Wave modelling - the state of the art. - Progress in oceanography, 2007. V. 75, # 4. P. 603-674.
8. Fomin, VN, Cherkesov, LV Simulation of drift currents in a shallow basin, taking into account changes in shear stress due to wind waves, Izvestiya, Ser. FAO. 2006. T. 42, № 3. p. 393-402.
9. Ardhuin F., Chapron B., Elfouhaily T. Waves and air-sea momentum budget: Implication for ocean circulation modeling / / J. Phys. Oceanogr. 2004. V. 34, # 7. P. 1741-1755.
10. Kabatchenko IM, Matuszewski G., Reznikov, M., Zaslavsky MM Simulation of wind and waves at the secondary thermal cyclones in the Black Sea. - Meteorology and Hydrology, 2001, N 5, p. 61-71.

11. Monin, AS Hydrodynamics of the atmosphere, ocean and Earth's interior .- Gidrometeoizdat, St. Petersburg, 1999, 524.

12. Monin AS, Yaglom AM Statistical Fluid Mechanics: Mechanics of Turbulence, v. 1. The MIT Press, Cambridge, Massachusetts, and London. 1971.

13. Zaslavsky MM On the parametric description of the drive layer of the atmosphere. - Math. Branch. Atmospheric and Oceanic Physics. 1995, t. 31, № 5, pp. 607-615.

14. Yan L. An improved wind input source term for third generation ocean wave modelling. Rep. No. 87-8. 1987: Royal Dutch Meteor. Inst .. 20pp.

15. Donelan M.A. et al., On the dependence of sea surface roughness on wave development. / / J. Phys. Oceanogr., 1993, vol. 23, pp. 2143-2149.

16. Kitaigorodsky SA Application of the methods of similarity theory to describe turbulence in the upper ocean. - Math. Branch. Atmospheric and Oceanic Physics. 1998, t. 34, № 3, pp. 430-434.

17. Soskin, IM Empirical relations to calculate the wind currents. - Proceedings of the SOI. Vol. 70, 1962, pp. 3-27.

18. Mastenbrock C., Burgers G., Jansen P. The dynamical coupling of a wave model and storm surge model through the atmospheric boundary layer .- J. Phys. Ocean. 1993. V. 23. № 8 P.1856 - 1866.